**Transport in hybrid electronic devices based on a modified DNA nucleoside (deoxyguanosine)**

*R. Rinaldi, E. Branca, R. Cingolani*

Istituto Nazionale per la Fisica della Materia (INFM)

and Dipartimento di Ingegneria dell'Innovazione, Università di Lecce, Italy

*R. Di Felice, E. Molinari*

INFM and Dipartimento di Fisica, Università di Modena, Italy

*S. Masiero, G.P. Spada, G. Gottarelli*

Dipartimento di Chimica Organica "A. Mangini", Università di Bologna, Italy

*A. Garbesi*

CNR IcoCea, Area della Ricerca di Bologna, Italy

Abstract

We report on a new class of hybrid electronic devices based on a DNA nucleoside (deoxyguanosine lipophilic derivative) whose assembled polymeric ribbons interconnect a submicron metallic gate. The device exhibits large conductivity at room temperature, rectifying behaviour and strong current-voltage hysteresis. The transport mechanism through the molecules is investigated by comparing films with different self-assembling morphology. We found that the main transport mechanism is connected to π-π interactions between guanosine molecules in adjacent ribbons, consistently with the results of our first-principles calculations.

Hybrid molecular-electronics has opened up the way to a new class of devices where the ultimate size reduction and reproducibility is provided by biological molecules interconnected to a multi-gated metallic nanopattern [1]. In principle, this approach has a twofold advantage: *(i)* The reduction of size down to the molecular level, not easily achievable by lithographic methods applied to inorganic materials, and, *(ii)* the high reproducibility of the device, which exploits the natural self-assembling capability of the biological species. The interconnection to the external world (power supply, load etc.) is on the other hand guaranteed by the conventional metallic part of the device. In this context, the identification of biological molecules capable of electrical conductivity, and the understanding of the dominant transport mechanisms, become crucial issues for the future exploitation of such hybrid systems. A possible route in this field is the use of DNA molecules to implement functional nanocircuits, which is particularly appealing given the possible implications of recognition mechanisms for fabrication [2-4]. However, the actual magnitude of DNA conductivity as well as its physical mechanisms are still largely debated, partially because of the very different conditions and measurements of the experiments [5-11]. A key issue is whether conduction takes place by ionic transport along the strands or outer mantle, or by electronic transport through the bases in the core. In the latter case, the basic mechanisms that are invoked include *(a)* electron transfer through direct tunneling (usually between acceptor and donor groups intentionally placed at a given distance along the DNA double–helix); *(b)* electron transport by sequential hopping or diffusion through localized states, similar to conduction in disordered solids; *(c)* band-like electron conduction through extended orbitals along the base stack. The latter mechanism is similar to band transport in semiconductors and virtually distance-independent, but would require a strong overlap between $\pi$ orbitals of adjacent bases in the stack.

In this Letter we address the problem of conductivity in hybrid organic/inorganic systems by fabricating a simple two-terminal device, consisting of a submicron metallic gate connected by a dried solution of lipophilic 3',5'-di-O-decanoyl-2'-deoxyguanosine (I), a modified DNA nucleoside (Fig. 1a). This molecule seems particularly attractive, because guanosine has the lowest oxidation potential among DNA bases [12]; furthermore, it spontaneously forms ribbon-like assembled species in solution (Fig.1a) [13] in the absence of ions, differently from natural polynucleotides which necessarily contain ionic species [2,4,5]. We then investigate electrical conductivity by means of current-voltage (I-V) measurements, and show that the system exhibits a non-ohmic behavior at room temperature, with strong hysteresis effects in the current-voltage characteristics. By comparing two structural configurations of the deposited layer corresponding to very different transport properties, we are able to extract information on the main microscopic mechanisms leading to conduction and we discuss them with the help of detailed ab-initio calculations performed for model guanosine-based solids.

The simple hybrid device consists of a metallic gate of width ranging between 300 and 800 nm is opened by Atomic Force Lithography in a Au/Cr (100nm/8nm) film deposited on glass. The deoxyguanosine molecules [13] are prepared in a $CHCl_3$ solution at low concentration (between $10^{-1}$ and $10^{-4}$ M) and deposited in the gate in the form of drops of constant volume (2 μl). The solution is then gently dried in vacuum until a solid film of deoxyguanosine molecules remains deposited in the gate. The morphology of the resulting film is studied in detail by Atomic Force Microscopy (AFM) experiments (using MMAFM Digital Nanoscope E). The I-V curves are recorded by applying the voltage with a signal generator and reading the current by means of an electrometer (noise level $<10^{-12}$ A).

The AFM topography of the molecular layers deposited in the gate, after complete drying of the solution, shows striking self assembling properties at very low densities. Fig.1b shows a regular arrangement of ribbons obtained with a single drop of gaunosine solution of concentration equal to $5*10^{-4}$ M. With increasing the number of molecules in the gate ( 3 drops in Fig.1c), the ordered morphology breaks down and a totally disordered film forms in the device. Such a morphology change influences dramatically the transport properties of the device: *in the ordered regime (Fig.1b) no measurable conductivity is found, whereas measurable conductivity is found in the disordered regime (Fig.1c) at suitable molecular concentrations*. This is exemplified in Fig. 2, where we show the measured I-V characteristics as a function of the number of drops deposited between the gold electrodes. In this experiment the concentration of the solution was fixed at about $10^{-1}$ M (disordered regime), i.e. about $10^{17}$ molecules/drop (from a simple geometrical analysis we estimate about $10^{10}$ molecules per drop in the gate). Under this condition the current increases from the noise level with one drop of solution up to the sub-μA range with 15 drops, and saturates at typical values of the order of 1 μA. The I-V curves are slightly asymmetric with respect to the bias, and exhibit an equivalent static resistance of the order 18 MΩ for the case of filled gap. This value compares to that recently reported for carbon-nanotube four-probe devices [14]. The asymmetry of the measured I-V curves can be related to capacitive effects due to polarization of the molecular layer.

We have also investigated the stability of the system after several I-V cycles in the range –10÷10 V. Strong hysteresis is seen by comparing the upward (open symbols) and downward sweep (solid symbols) in Fig. 2, indicating some charge accumulation in the organic film, probably originated by the additional capacity introduced by the guanosine film in the metallic device. For high driving voltages (above 10 V) the current is quite large, causing the degradation of the I-V curves (increase of device resistance) with increasing the number of sweeps. However, for sweeps below 6 V the I-V curve does not show substantial degradation after many cycles at room temperature.

We now discuss the origin of the observed transport properties. We first point out that ionic conduction mechanisms can be ruled out, as immediately inferred by the non-linear dependence of the current on the number of drops at a fixed bias, as opposed to the linear dependence expected for ionic conductivity. On the other hand, the transport properties of the device depend crucially on the arrangement of the ribbons in the film as observed by AFM. In the self-assembled ordered phase the orientation of ribbons is incompatible with a π-π stacking of several molecules belonging to adjacent ribbons, with the stacking direction connecting the electrodes. The insulating behavior observed in the ordered phase implies that charge transport cannot take place along the ribbons, i.e. it does not occur within a surface containing the guanosine molecules.

On the other hand, in the disordered phase the film does not display any macroscopic ordering parallel to the substrate. With increasing the number of drops deposited in the gate, the distribution of ribbons is randomized, and the molecules are mostly oriented with their planes *not* parallel to the substrate. Therefore, the probability increases to obtain local configurations with high π-π overlap among molecules in adjacent ribbons. The high conductivity measured in this regime therefore suggests that transport is related to *charge transfer between different ribbons, occurring in the direction perpendicular to the surface containing the molecules*.

The above experimental results point to a conduction mechanism where electron delocalization due to π-π overlap between guanosine molecules plays an important role, at least within some local regions which might then be connected via tunneling or hopping processes.

To further examine this possibility, we have investigated the occurrence of a band-like transport mechanism for stacked molecules along their stacking direction by a first-principle study of model periodic solids. Our calculations are based on the Density Functional Theory (DFT) in the Local Density Approximation (LDA) [15].

The molecule-molecule interactions are investigated by assuming different periodic arrangements of guanosines along the *z* direction perpendicular to the *xy* molecular planes. The *z* stacking distance was fixed at the calculated value of 3.37 Å (in good agreement with the X-ray determination [13]), while the molecules were isolated in the *xy* planes. We have examined a very large number of configurations. For each of them, we have relaxed the atomic positions until the forces vanished, consistently with the full quantum mechanical electronic structure [16]. Our results are exemplified in Fig.3 for two typical geometries with two guanosine molecules in the unit cell. The formation of the periodically stacked solid, with respect to the isolated molecules, is endothermic by 100 meV/molecule for configuration A (Fig.3a), and exothermic by 280 meV/molecule for configuration B (Fig.3b). From an accurate analysis of the electron states, we attribute the higher formation energy of configuration A to the π-π repulsion between the Highest Occupied Molecular Orbitals (HOMO) lying on top of each other. For configuration B, such a repulsion is reduced by the in-plane shift between neighboring π orbitals.

The filled and empty bands are found to be separated by a LDA energy gap of 3.0 eV (A, see Fig. 3c) and 3.6 eV (B, see Fig. 3d): these values are affected by the typical LDA underestimate of about 50%. As a consequence of the π–π interaction, the energy bands derived from the HOMO, as well as those derived from the Lowest Unoccupied Molecular Orbital (LUMO), are found to be significantly dispersive (Fig.3). The calculated hole ($m_h$) and electron ($m_e$) effective masses, in units of the free electron mass, are $m_h = 1.0$ and $m_e = 1.4$ for configuration A, $m_h = 2.2$ and $m_e = 2.8$ for configuration B. These values are larger with respect to conductive polymers by an order of magnitude [17], but lie in a range characteristic of heavy carriers in wide-band-gap semiconductors [18]. The largest is the superposition of π orbitals, the lowest are the effective masses, for which better transport performances are expected.

It is important to stress that, even if a larger π-π superposition tends to reduce the structural stability, the energy differences that we have found are very small. This suggests that a combination of possible stacking arrangements may occur in nature, exhibiting the conditions for band transport of holes in the valence band or electrons in the conduction band as soon as the π orbitals show significant superposition. We therefore propose that the films where conduction was observed (disordered regime) be comprised of local regions where sufficient π–π interaction occurs to yield extended wavefunctions and band-like conduction; these regions will then be connected, probably by incoherent tunneling or hopping processes, to give overall charge transport across the electrodes. Conversely, the absence of conduction in the ordered samples can be attributed to their geometrical configuration that inhibits π–π interactions.

Finally, it is worth mentioning that the semiconducting band structure resulting from our calculations indicates the possibility that transport properties could be further improved by increasing the carrier density in the system through intentional doping. Preliminary experiments performed on similar samples that contained traces of Na$^+$ and K$^+$ ions indeed show the same regimes concerning the film morphology, with strongly increased conduction in the disordered regime.

In conclusion, we have fabricated simple two-terminal devices consisting of an Au/Cr gate connected by a film of deoxyguanosine molecules. When the film surface shows no apparent ordering, the I-V curves exhibit strong conductivity, a slightly asymmetric rectification, strong hysteresis cycle and modest aging for bias voltages below 6 V at room temperature. On the other hand, when the molecular ribbons are self-ordered parallel to the substrate, no current transport is observed. Based on first-principles band-structure calculations for prototype guanosine-based solids, the result are interpreted as the consequence of the very different π-π overlap between guanosine molecules occurring in the two configurations.

Useful discussions and collaboration with G. Proni in the early stages of this work are gratefully acknowledged, together with the expert help of M. De Vittorio and P. Visconti. This work is supported in part by MURST 40%.

References

[1]  J. Jortner and M.A. Ratner (eds.), *Molecular Electronics* (Blackwell, London, 1997).

[2]  D. Porath et al., Nature **403**, 635 (2000).

[3]  E. Braun et al., Nature **391**, 775 (1998).

[4]  Y. Okahata et al., J.Am.Chem.Soc. **120**, 6165 (1998).

[5]  H.-W. Fink and C. Schonenberger, Nature **398**, 407 (1999).

[6]  J. Warman, M.P. de Haas, and A. Rupprecht, Chem. Phys. Lett. **249**, 319 (1996).

[7]  M.G. Debije, M.T. Milano, W.A. Bernhard, Angew. Chem. Int. Ed. **38**, 2752 (1999).

[8]  M. Bixon et al., Proc. Nat. Acad. Sci. **96**, 11713 (1999).

[9]  F.D. Lewis et al., Science **277**, 673 (1997).

[10] S.O. Kelley and J.K. Barton, Science **283**, 375 (1999).

[11] B. Giese et al., Angew. Chem. Int. Ed. **38**, 996 (1999).

[12] D. Ly, L. Sanii, and G.B. Shuster, J. Am. Chem. Soc. **120**, 9400 (1998), and references therein.

[13] G. Gottarelli et al., Helv. Chimica Acta **81**, 2078 (1998), and, G. Gottarelli et al., Chem. Eur. J.

(2000), in press.

[14] A. Bezryadin et al., Phys. Rev. Lett. **80**, 4036 (1998).

[15] R.M. Dreizler and E.K.U. Gross, "Density Functional Theory. An Approach to the quantum many-body problem.", Springer-Verlag, Berlin (1990).

[16] M. Bockstedte et al., Comp. Phys. Comm. **107**, 187 (1997).

[17] see e.g. P. Gomes Da Costa and E.M. Conwell, Phys. Rev. B **48**, 1993 (1993).

[18] S.K. Pugh et al., Semicond. Sci. Technol. **14**, 23 (1999).

FIGURE CAPTIONS.

Fig. 1

*(a)* Schematics of a single deoxyguanosine molecule (top) and sketch of the ribbon-like structure formed by the deoxyguanosine molecules connected through hydrogen bonds (bottom). Here R is a radical containing the sugar and alkyl chains (see Ref. 13).

*(b)* AFM micrograph of the ordered solid state deoxyguanosine film obtained after drying one drop of the solution in the gate (with concentration of approximately $5 \cdot 10^{-4}$ M). The regular arrangement of ribbons has a lateral periodicity of about 4 nm. The ribbon width of about 3 nm is consistent with that determined by X-ray measurements. The image could be interpreted as due to a regular array of ribbons with the heterocyclic moiety on the glass substrate. It is worth noting that the self-assembling geometry observed by AFM depends very sensitively on the substrates: deposition under the same conditions on the gold electrodes rather than in the glass gap causes the formation of parallel ribbons spaced by about 2.5 nm, with no indication of coiling.

*(c)* AFM micrograph of the solid state deoxyguanosine film obtained after drying three drops of the same solution in the gate: the ordered morphology in this case is lost. All the AFM measurements were performed with atomic resolution in contact mode on the top surface of the film.

Fig. 2

Current-voltage characteristics obtained at room temperature, with increasing the number of drops in the gate (note that the absolute values of the current are plotted on a log scale). The concentration of the solution was about $10^{-1}$ M. Open symbols refer to currents measured while increasing the gate voltage from –10 to +10 V, solid symbols refer to currents measured while decreasing the voltage from +10 to –10 V. The noise level of the set-up is shown by the bottom curve (dotted line), measured through the gate in the absence of the deoxyguanosine layer.

Fig. 3

Top view of two optimized geometries and corresponding band structures. *(a)* Configuration A: the two molecules in the unit cell lie on top of each other, in such a way that the HOMO and LUMO π orbitals are exactly superposed. *(b)* Configuration B: the upper molecule (dashed lines) and the lower molecule (solid lines) are rotated and shifted: in the optimized geometry, the HOMO and LUMO π orbitals are not on top of each other, but laterally shifted. *(c)* and *(d)* Calculated band dispersions along the *z* direction for configurations A and B, respectively, with the HOMO and LUMO bands indicated with larger dots. The calculated electron energy eigenvalues are plotted against *k* vectors perpendicular to the *xy* planes of the molecules, between the center (Γ) and edge (A) of the Brillouin Zone.

Fig.1a)

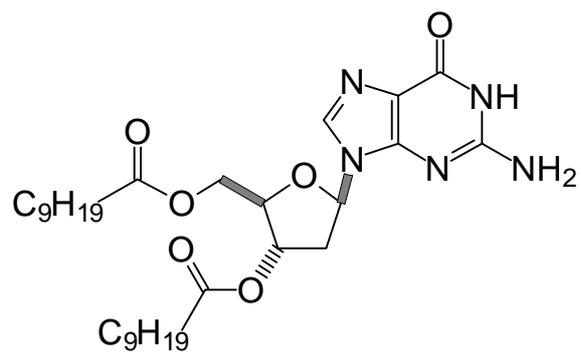

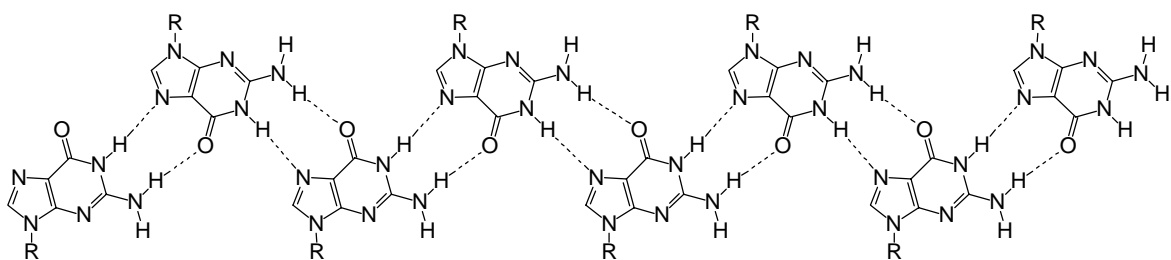

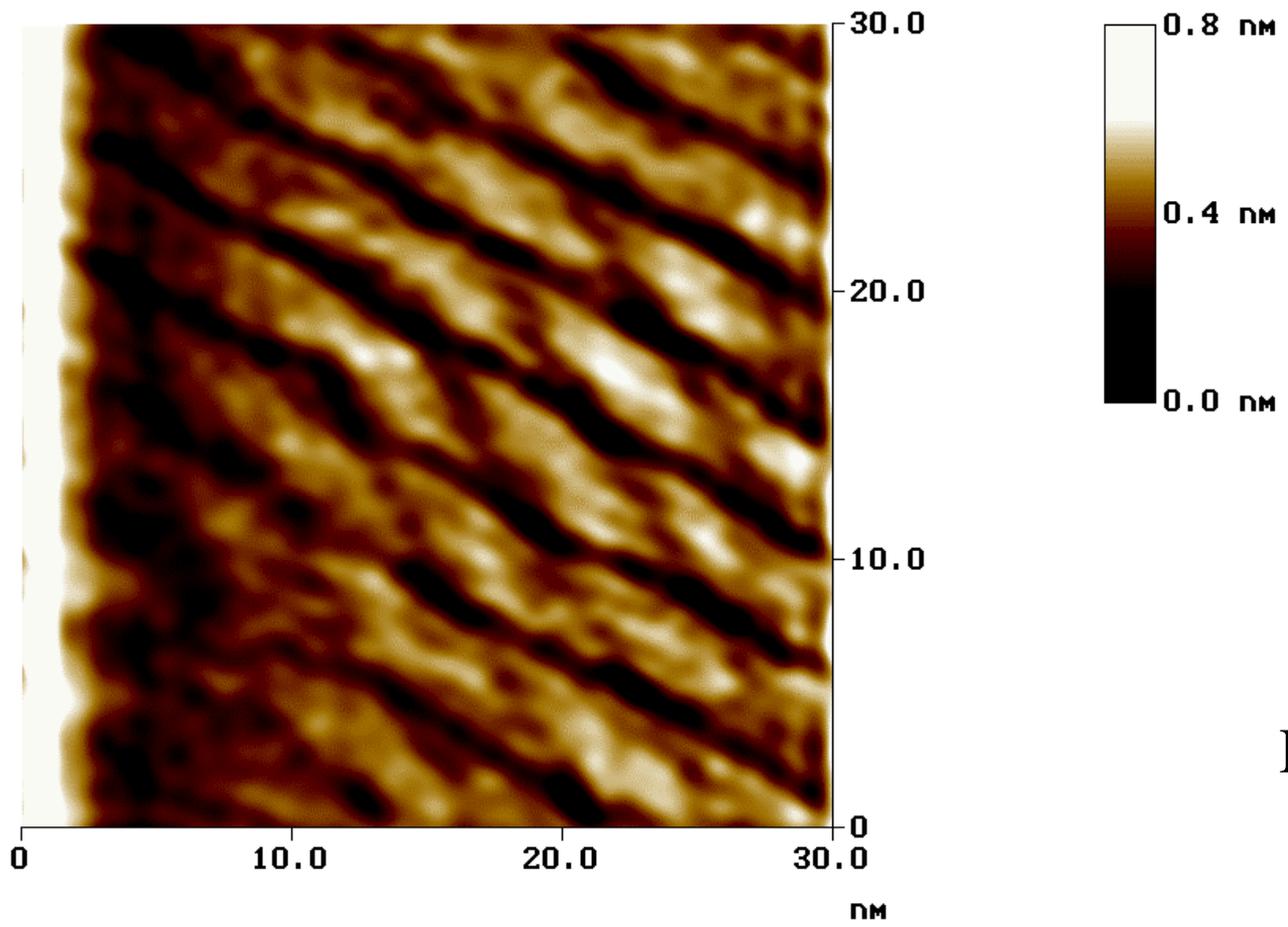

Fig.1b

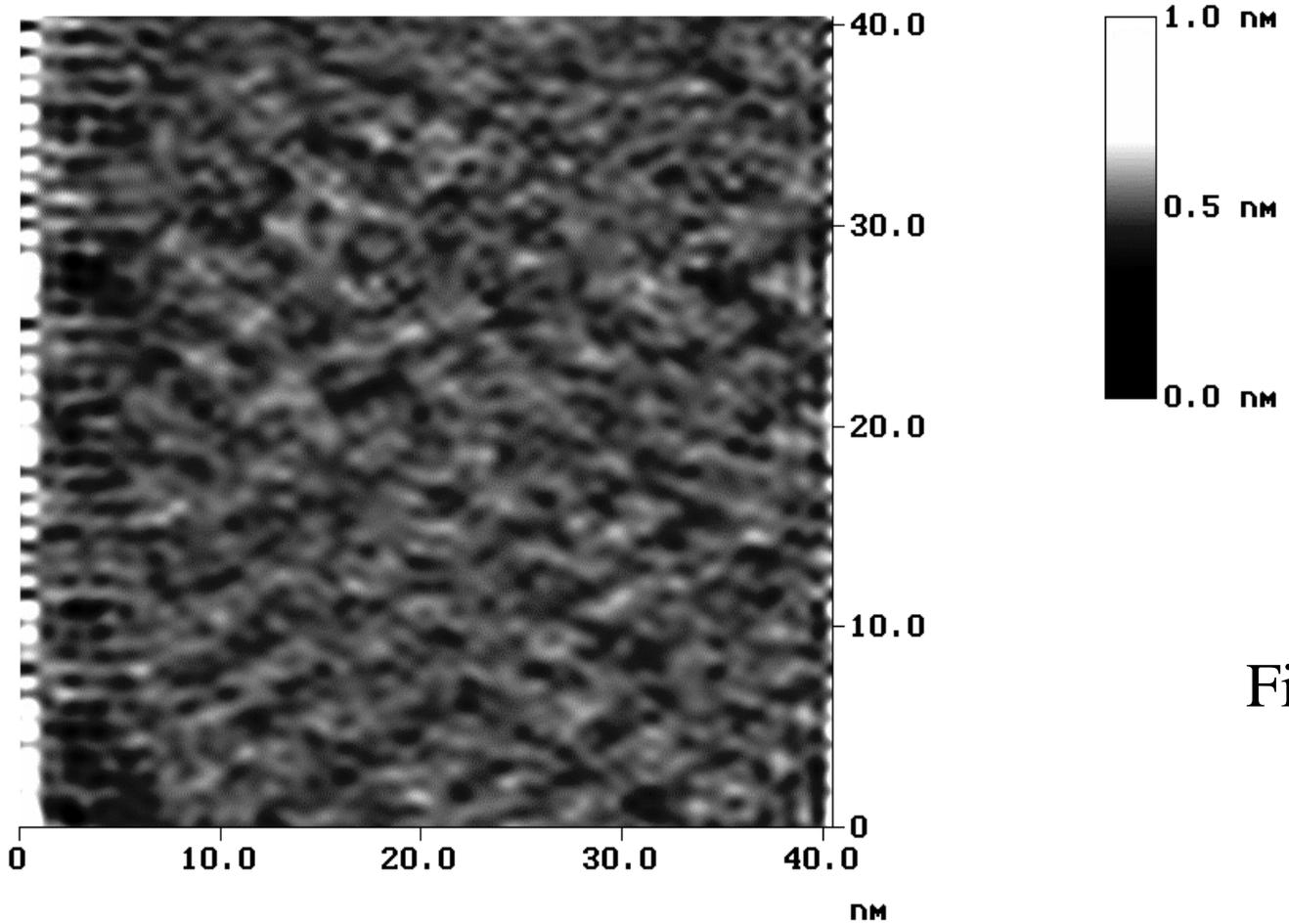

Fig.1c)

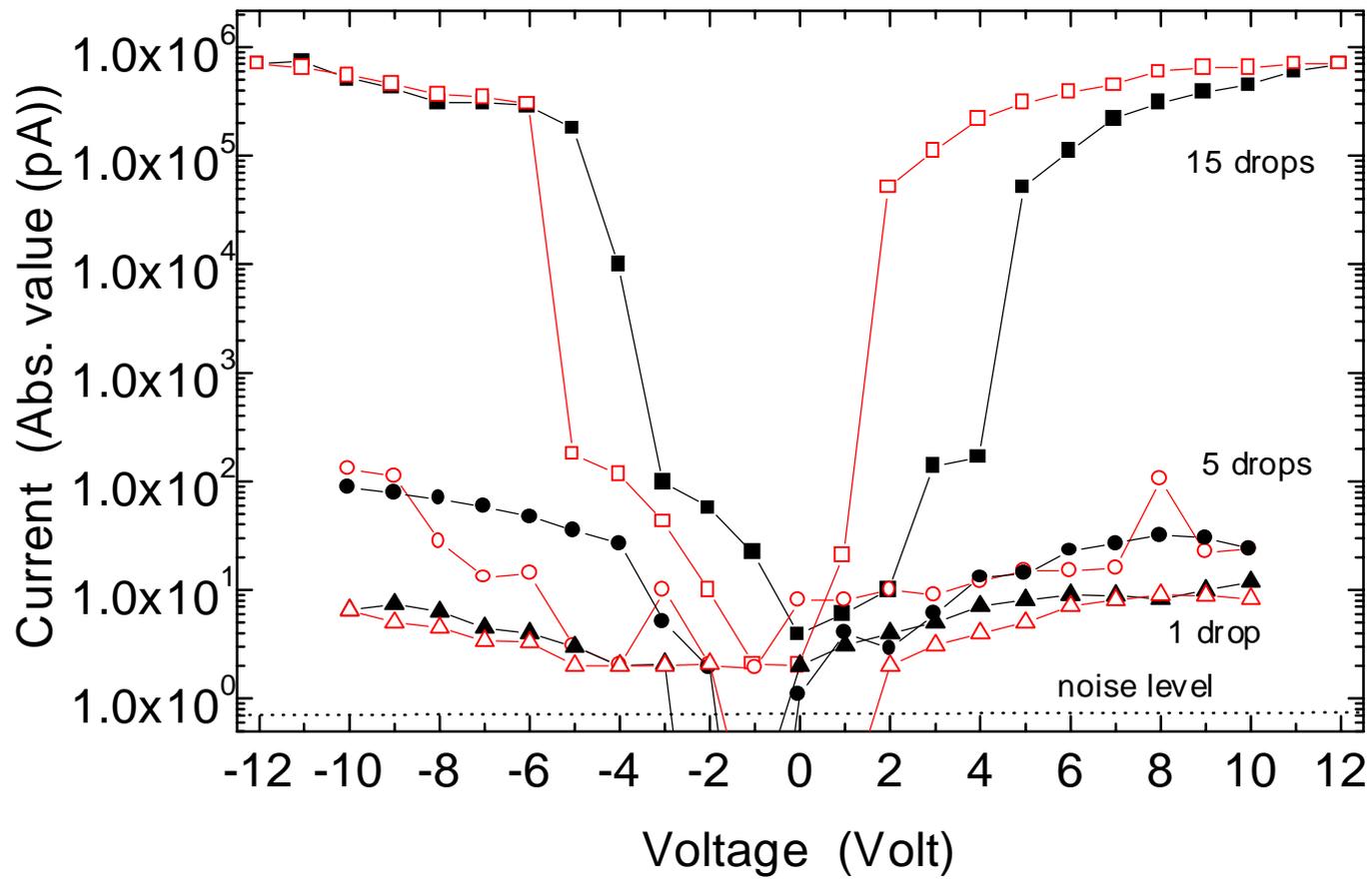

Fig.2

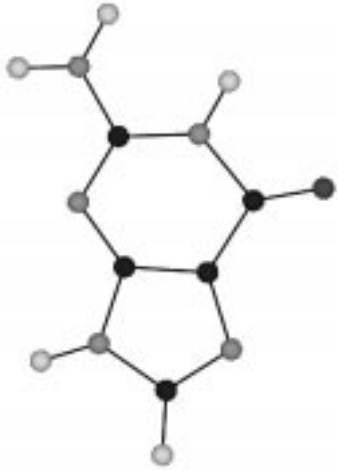 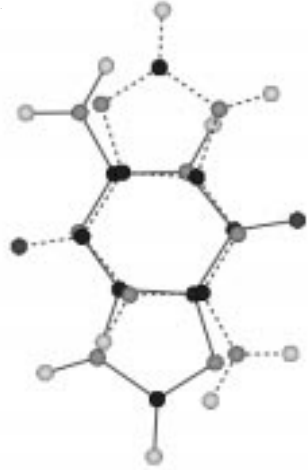

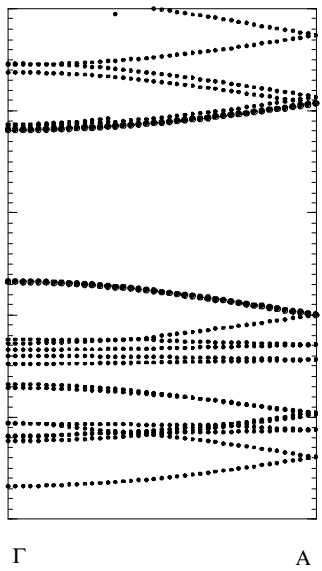 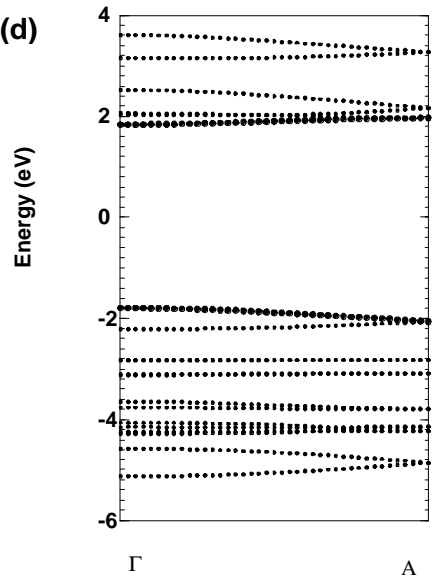

Fig.3